\documentclass[runningheads]{llncs}
\usepackage[T1]{fontenc}
\usepackage{graphicx}
\usepackage{here}
\usepackage{newunicodechar}
\newunicodechar{’}{'}

\begin{document}
\title{On-Demand Lecture Watching System \\Using Various Actions of Student Characters \\to Maintain Concentration}
\titlerunning{On-Demand Lecture System Using Various Actions of Student Characters }
%
\author{
Saizo Aoyagi\inst{1}\orcidID{0000-0003-2359-1109} \and
Ryoma Okazaki\inst{2} \and
Seishiro Hara\inst{2}  \and \\ 
Fumiya Ikeda\inst{3} \and
Michiya Yamamoto\inst{3} 
}

\authorrunning{A. Aoyagi et al.}
%
\institute{Faculty of Global Media Studies, Komazawa University, Setagaya-ku Tokyo 154-8525, Japan 
\email{aoyagi12@komazawa-u.ac.jp}
\and
Graduate School of Science and Technology, Kwansei Gakuin University, Sanda, Hyogo 669-1330, Japan
\and
School of Engineering, Kwansei Gakuin University, Sanda, Hyogo 669-1330, Japan
}
\maketitle              
\begin{abstract}
Since the COVID-19 pandemic, online lectures have spread rapidly and many students are satisfied with them.
However, one challenge remains the loss of concentration due to the lack of students' copresence.
Our previous work suggests that presenting 3D characters with appropriate actions has the potential to improve concentration in online lectures.
Nevertheless, an effective combination of actions has not yet been identified.
In this study, we developed a lecture watching system that presents a 3D virtual classroom using a naked-eye 3D display. 
The system includes student characters that show copresence with various actions such as nodding, notetaking, and sleeping.
An evaluation experiment was conducted with two conditions; (1) student characters perform only positive actions and (2) both positive and negative actions.
The results, analyzed using posture and notetaking behavior as key indicators, suggest that the system can help to maintain concentration when the student characters perform both positive and negative actions, rather than only positive ones.
These findings provide promising strategies for maintaining student focus in on-demand lectures and contribute to the development of more effective online education systems.
\end{abstract}
\section{Introduction}
\footnote{This is the author's final draft}
During the COVID-19 pandemic, online lectures increased rapidly. 
Many institutions still offer them, often combined with face-to-face classes in a HyFlex design, due to advantages such as attending from home and replaying recorded videos at any time. 
Lee et al. \cite{オンライン満足度} reported high satisfaction ratings for online lectures, which confirms these benefits. 
With advances in information technologies, educators have also begun adopting new technologies such as virtual reality (VR). 
For example, N High School in Japan uses VR classes, allowing students to hold class materials such as graphs and molecular structures in their hands \cite{N高}.

Despite these benefits, online education still faces challenges, including reduced concentration and decreased motivation \cite{オンデマンド授業の影響}.
One major reason, as Fujii et al. \cite{キャラクタ関わり} noted, is the lack of copresence with other students.
In a classroom setting, students can engage in discussions and casual chats, and a communication field is activated that fosters further communication. 
Moreover, Sunaga pointed out the importance of social facilitation, where the mere presence of others can positively influence individual performance \cite{社会的促進}.
Also, Yang et al. found that the social presence of students in live lectures increases learning satisfaction \cite{社会的存在}. 
We hypothesize that the presence of classmates for students can improve the learning experience in on-demand lectures.

Embodiment in communication plays an important role in fostering a copresence. 
In face-to-face settings, nodding or gesturing can lead to physical entrainment, helping to create unity within the communication field shared by the speaker and listeners \cite{渡辺_身体性}. 
In our earlier work, overlaying two nodding student characters on a lecture video improved mini-test scores \cite{うなずくキャラクタ}. 
Subsequently, we increased the number of student characters and varied their nodding behaviors. 
The pattern in which the nodding and negative actions are alternated showed a temporary improvement in the audience retention rate compared to the pattern in which the nodding action is always performed.
However, the characters still appeared somewhat unnatural, which prevented a complete assessment of their variety of actions \cite{うなずき・}.
We then examined how to present these characters more naturally as a group and found that including negative actions, such as sleeping, rather than only positive ones, was important \cite{自然な集団学生キャラクタ}. 
However, for longer videos, which are usually used for on-demand lectures, it is still unclear how to control the range of character actions to further increase engagement or maintain learners' concentration.

Therefore, in this study, we developed an on-demand lecture watching system with multiple student characters whose different actions maintain concentration of students. 
The contribution of this study is providing promising a strategy for maintaining student focus in on-demand lectures with action modes that alternate positive and negative behaviors among virtual students, reflecting realistic classroom dynamics, and maintaining learner concentration over extended viewing periods.

\section{Related Studies}

\subsection{Copresence on Real and Online Classes}

Face-to-face lectures clearly demonstrate the importance of other students' presence and collaboration. 
Okada (2008) suggests that learning with friends increases students' sense of fulfillment \cite{岡田涼}. 
Furthermore, Okada et al. (2017) found that collaborating with others facilitated knowledge gains in adult learners \cite{岡田晋作}. 
In our own survey, roughly a third of the respondents reported working mainly with friends in class 
 \cite{広瀬}. 
These findings underscore the importance of learning together in a shared environment, and we believe that this principle is applicable to online or virtual lectures. 

Online lectures typically take two forms: 
(1) synchronous and real-time lectures using platforms like Zoom and on-demand lectures, where students watch recorded lectures on their own schedule.
In synchronous lectures, when everyone turns on their camera, students and the instructor can see each other's faces and reactions, although it remains less interactive than a physical classroom. 
(2) In contrast, on-demand lectures flow mostly in one direction, from the instructor to students, which can decrease learner motivation due to limited real-time feedback. 
Although some on-demand lectures incorporate comment sheets or feedback after each session, they provide few opportunities to feel copresence with classmates while watching.

Various attempts have already been made to incorporate computer-ganerated (CG) characters into educational platforms as virtual peers. 
For example, in VR lectures at N High School, for example, it is possible to record student movements and play them back to other learners using characters on the screen \cite{N高}.
In another project, Zivi et al. built a GPT-4-based agent to highlight key parts of the content or to prompt learners to focus, encouraging more active learning \cite{ClassMeta}.
Watanabe et al. \cite{渡辺} placed an agent on student computers to provide advice and explain terminology, finding that students' motivation improved when they sensed they were learning alongside that agent. 
Mayer et al. \cite{Mayer} showed that an agent that moves in a human-like manner leads to higher test scores than no agent at all. 
Although these solutions can make characters more realistic, overly conspicuous character actions could end up distracting students from essential lecture content.
We also replicate other learners in place of real classmates for the purpose of creating a communication field and copresence and maintaining concentration, which is different from other researches.

\subsection{Generation of Communication-Field}
The existence of other people is crucial in creating a communication field in many other fields. 
For example, Oshiro et al. created a method to synchronize the cheering movements of a virtual audience with the actions of users during live music, enhancing a sense of unity with that virtual crowd \cite{ペンライト}. 
Ogawa et al. reported that adding avatars to online concerts boosted participants' involvement  and unity with one another \cite{ライブ_アバタ}. 
Likewise, Sejima et al. demonstrated that a virtual audience that reacts physically in response to spoken content can make communication more active \cite{場の盛り上がり}.
We believe that such a simulated peer presence could also increase students' motivation or concentration of during lectures.

\subsection{XR for Online Education}


Driven by government initiatives and the wider adoption of online learning during the pandemic, schools and universities now often equip each student with a PC. 
Meanwhile, xR technology has advanced to the point that some schools have implemented VR-based lessons. 
For instance, N High School relies on VR headsets to immerse students in a virtual classroom \cite{N高}. 
Although this strategy can deliver a strong sense of presence and better 3D visualization (e. g., for molecular structures), VR headsets can cause ``VR sickness'', feel constrictive, or appear bulky when wearing for long periods.

Naked-eye 3D displays, on the other hand, offer stereoscopic visuals without a headset, typically imposing less fatigue. 
It is also called autostereoscopic display, and various technologies realize this feature.
Lenticular-based autostereoscopic displays are promising and are used in the fields of digital signage, game display, education, CAD, and product design \cite{autostereoscopic}. 
This technology enables users to continue taking notes or referring to other materials while viewing with naked-eye 3D displays in on-demand lectures. 
This capability makes it feasible to continue to watch an extended lecture without undue discomfort.



\vspace{-0.2cm}

\subsection{Approach of This Study}
Before constructing our main system, we conducted a preliminary experiment to clarify how to make the group of student characters move naturally.
Participants evaluated factors such as how the timing of each action change, the number of distinct action types, and the placement of negative-action characters (e.g., sleeping) influenced perceived naturalness. 
The results indicated following: (1) more varied actions appeared more natural; (2) assigning negative-action characters to classroom's rear looked more realistic while keeping a generally positive impression; and (3) even if the timing is slightly off, it feels natural.
We incorporated these insights into seating layout and character actions.

 \vspace{-0.2cm}

\section{Concept of the System}
\begin{figure}[bth]
		\begin{center}
		\hspace{-0.3cm}
        \includegraphics[width=0.6 \textwidth]{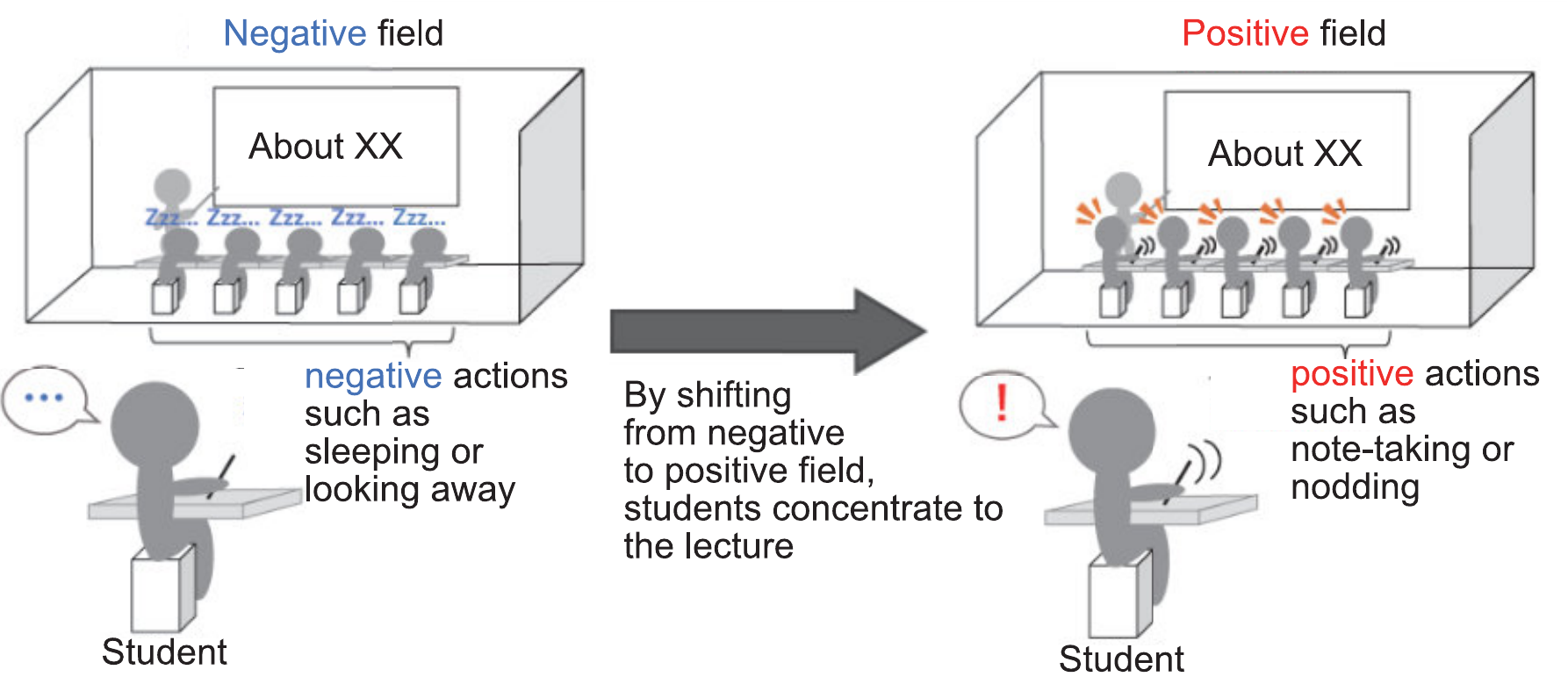}
		\end{center}
		\vspace{-0.6cm}
		\caption{Concept of the system.}
		\label{fig:コンセプト図}
\end{figure}

\vspace{-0.2cm}
Our on-demand lecture watching system inserts a group of student characters, each performing various actions, into a CG classroom to create copresence and foster a communication field. 
By controlling the characters' actions, we can shape this field to maintain students' concentration. 
For example, having them nod or take notes can encourage a sense of active learning, and occasionally introducing negative actions, such as looking away, can help raise interest when switching back to positive actions (Fig. \ref{fig:コンセプト図}).

\vspace{-0.4cm}

\vspace{-0.4cm}
\clearpage

\section{Development of the System}
\subsection{Overview of the System}

\begin{figure}[tbh]
		\begin{center}
			\includegraphics[width=0.9\textwidth]{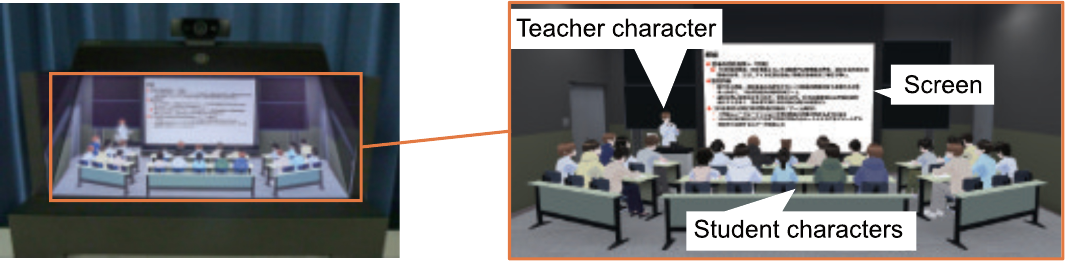}
		\end{center}
		\vspace{-0.4cm}
		\caption{Lecture-watching system.}
		\label{fig:開発した講義受講システム}
\end{figure}

Building on our previous system \cite{自然な集団学生キャラクタ}, we used Unity 2021.3.24f1 (Unity Technologies), along with a Sony ELF-SR1 naked-eye 3D display to develop a platform for on-demand lectures (Fig. \ref{fig:開発した講義受講システム}). 
In the virtual classroom, we placed a large screen for the lecture video, an instructor character, and 22 student characters. 
Each student character can be in one of the following states:

\begin{itemize}
  \item Positive actions: notetaking, nodding
  \item Negative actions: sleeping, looking away, leaning on an elbow
  \item Neutral actions: leaning forward, sitting upright
\end{itemize}

We used an Azure Kinect DK (Microsoft) to record these actions, syncing them with the lecture timeline. 
To avoid appearance of repetitiveness among the characters, each action has between one and three variations. For notetaking specifically, the character looks at the on-screen slides for 0.5 to 1.0 seconds, and then gazes at its notebook for 3.0 to 4.0 seconds.
We scheduled a sequence of actions (type and timing) for each character in a CSV file, making it easier to manage longer videos.

\subsection{Action Modes of Student Characters}\label{section:集団学生キャラクタの動作モード}

A previous study \cite{うなずき・} suggested that changing student characters' actions over time could promote viewing, although conclusive evidence was lacking. 
Thus, we designed two action modes for extended lectures: Stable Mode and Dynamic Mode.

\begin{figure}[tbh]
		\begin{center}
			\includegraphics[width= 1 \textwidth]{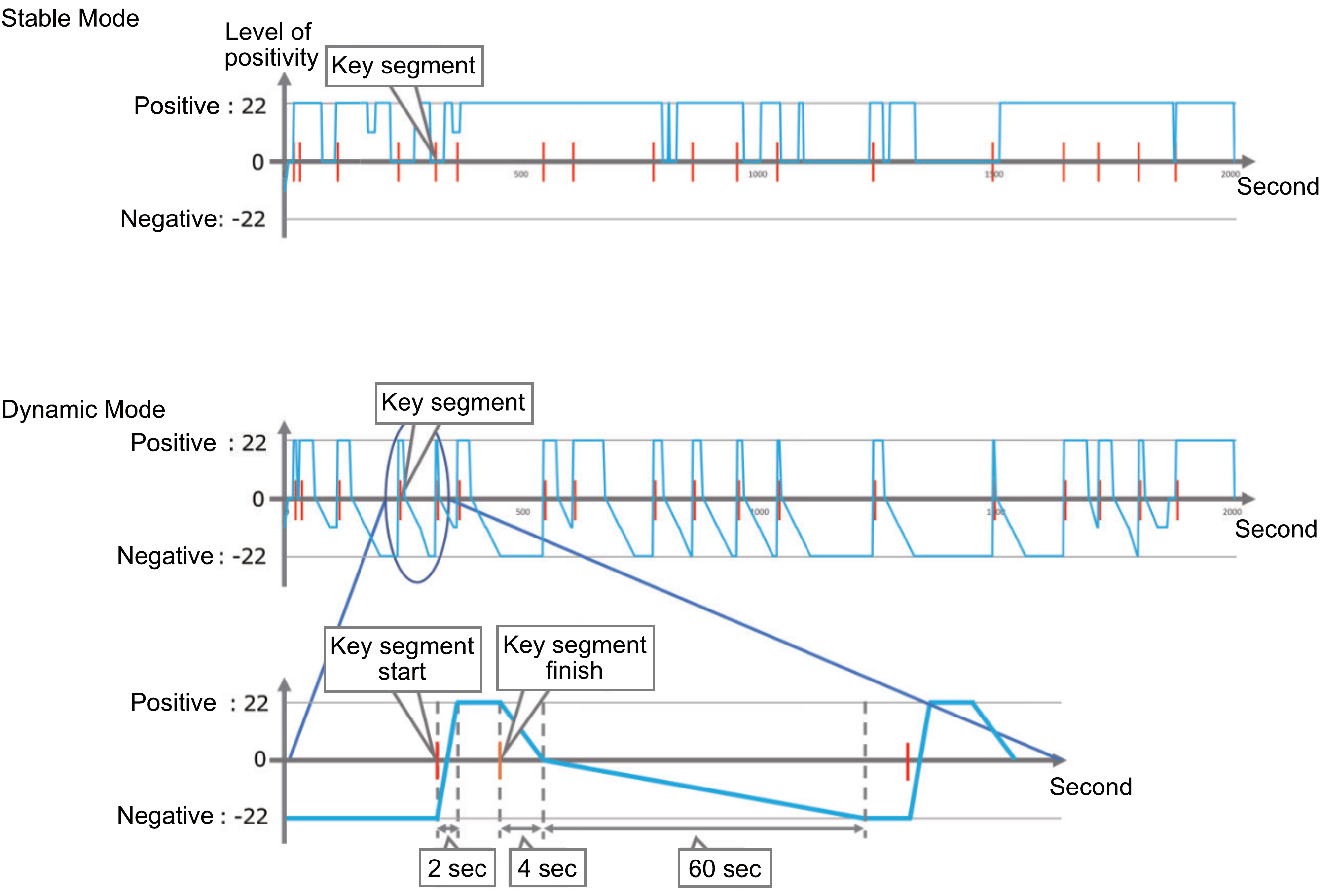}
		\end{center}
		\vspace{-0.3cm}
		\caption{Explanatory diagram of the two action modes.}
		\label{fig:動作モード_修論}
\end{figure}

Figure \ref{fig:動作モード_修論} shows the level of positivity among the groups of student characters in each mode.
There are 22 characters in each group.
For each character, we assign +1 when performing a positive action, -1 when performing a negative action, and 0 otherwise, and then we summed these values for the 22 characters. 
As a result, the total positivity of the group can range from +22 down to -22.

\hspace{1em}

\noindent{
\textbf{Stable Mode}
}

\begin{figure}[tbh]
		\begin{center}
			\includegraphics[width= 0.6 \linewidth]{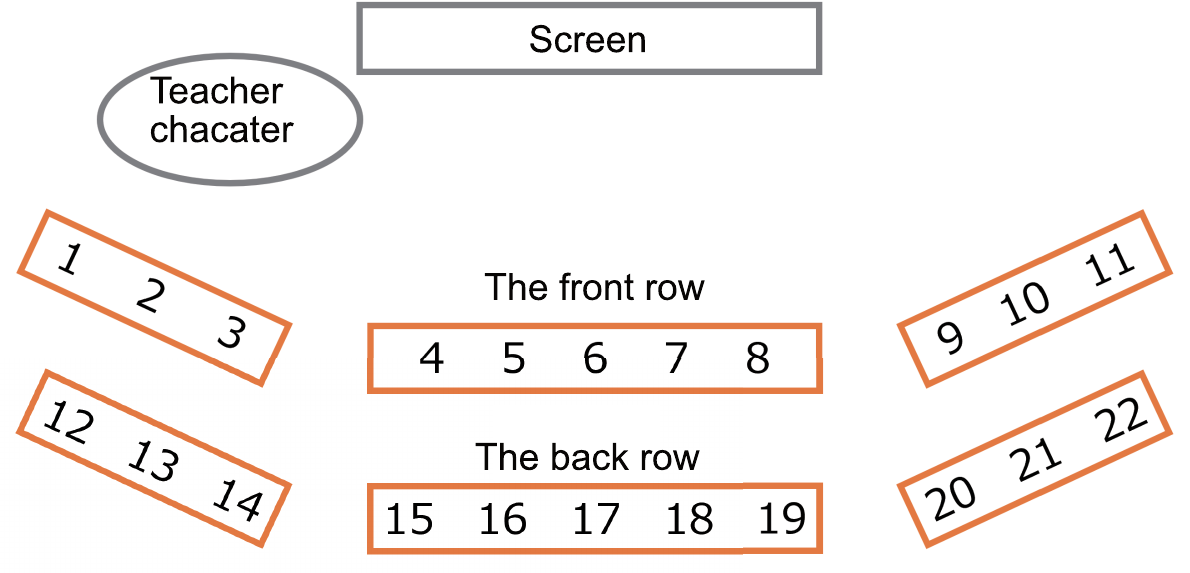}
		\end{center}
		\vspace{-0.3cm}
		\caption{Student character seating chart.}
		\label{fig:座席表}
\end{figure}

In this mode, nearly all student characters maintain positive actions for most of the lecture. 
Every 18-30 seconds, half of the front row (Fig. \ref{fig:座席表}) switches between notetaking and nodding, followed by a similar switch in the back row. 
While an embedded video segment is playing, the characters first switch to leaning forward or sitting upright, and then they randomly change actions at five equally spaced intervals.

Stable Mode fosters consistently positive actions that can encourage concentration.
On the other hand, less change in character movement may reduce students' concentration.

\vspace{-0.4cm}
\noindent{
\textbf{Dynamic Mode}
}

\noindent

In this mode, most student characters generally remain in negative actions but switch to positive ones shortly before and during key lecture segments. 
Starting two seconds before each key segment, they begin transitioning to positive actions, maintain them until the segment ends, and then gradually revert to negative actions. 
They all lean forward for 4 seconds, and then move from the far edges of the back row to the center of the front row over the course of 60 seconds, returning to negative actions. 
In total, the characters spent 24.8\% of the lecture time in positive actions, 42.1\% in transition, and 33.1\% in negative actions.

To make sure the actions began naturally in both modes, immediately after the lecture commenced, students in the front row were asked to lean forward and sit up straight, whereas those in the back row were asked to lie down, look away, or put their elbows on their desks, with each row performing these actions at random.
In addition, when transitioning between the actions of multiple characters, it was shown that shifting the timing would make it more natural, so the time it took for all characters to transition between actions was set at 2 to 4 seconds and the timing was shifted randomly within that time.
Dynamic Mode may be more effective in drawing viewers' attention.


%

\vspace{-0.2cm}

\section{Evaluation Experiment}

\vspace{-0.2cm}


We conducted an evaluation from December 12 to December 22, 2023, in the ``Computer Graphics'' course at a university's School of Engineering, with 50 undergraduate participants (39 men, 11 women, average age 20.8).

\begin{figure}[bth]
	\begin{center}
		\begin{tabular}{c}
		\hspace{-0.4cm}
			\begin{minipage}{0.5\hsize}
                \vspace{-0.35cm}
				\begin{center}
					\includegraphics[height= 0.35 \linewidth]{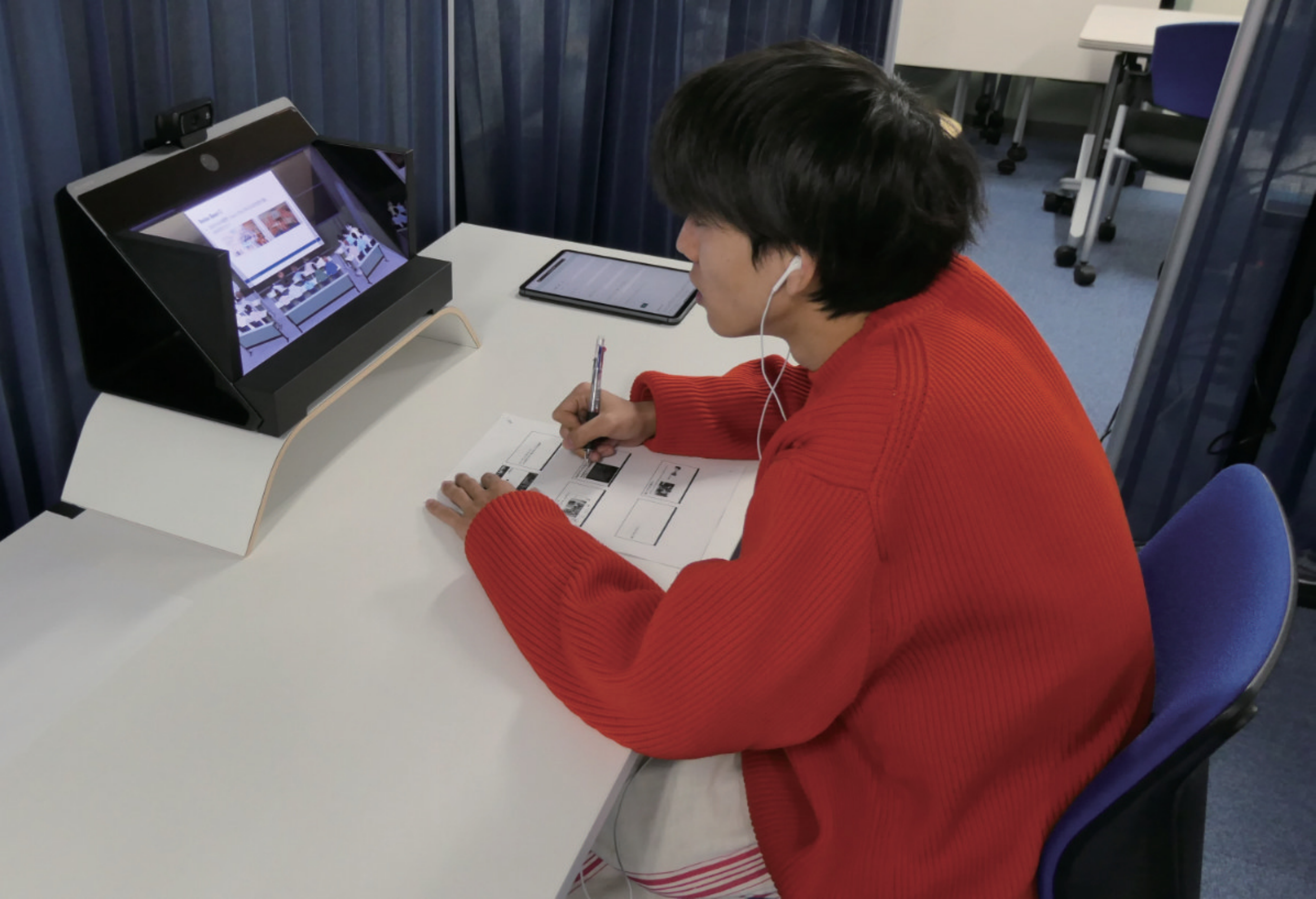}
				\end{center}
				\vspace{-0.7cm}
				\caption{Experimental scenery.}
				\label{fig:実験の様子_修論}
			\end{minipage}
			\begin{minipage}{0.5\hsize}
				\vspace{-0.35cm}
				\begin{center}
					\includegraphics[height= 0.35 \linewidth]{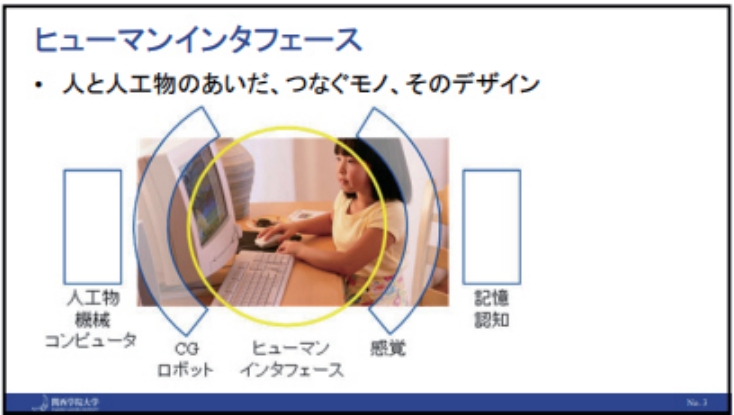}
				\end{center}
				\vspace{-0.7cm}
				\caption{An example of the slide.}
				\label{fig:スライドの例}
			\end{minipage}
		\end{tabular}
	\end{center}
\end{figure}

Because this experiment was conducted as a part of an actual lecture, we administered a mini-test at the end. 
Each student also received an A4 handout of the lecture slides (Fig. \ref{fig:スライドの例}) for notetaking. 
The 33-minute lecture, titled ``Computer Graphics Extra'', was a prerecorded session consisting of 29 slides. 
We generated 19 mini-test questions that focused on the key segments of the lecture. 
Participants were randomly assigned to Stable Mode or Dynamic Mode (25 each). 
For analysis, we filmed them from the left side of the students, clarifying that posture would not affect their grades.


The procedure was as follows:
(1) Participants watched the 33-minute video (Stable Mode for one group, Dynamic Mode for the other); 
(2) they had 3 minutes to review their notes, after which we collected the handouts.
(3) they answered a 19-question multiple-choice mini-test via Microsoft Forms (Fig. \ref{fig:実験の様子_修論}); and
(4) they answered the lecture impression questionnaire on the same platform.
\vspace{-0.4cm}

\clearpage

\subsection{Evaluation item:}
We examined four main factors.

\subsubsection{Questionnaire of impression:}

Participants rated 10 items on a 7-point Likert scale (1 = strongly disagree to 7 = strongly agree), including statements such as ``I felt a positive atmosphere'' and ``I felt the class changed the pace.''
The list of all items is shown in the next chapter.


\begin{figure}[H]
		\begin{center}
			\includegraphics[width= 0.6\linewidth]{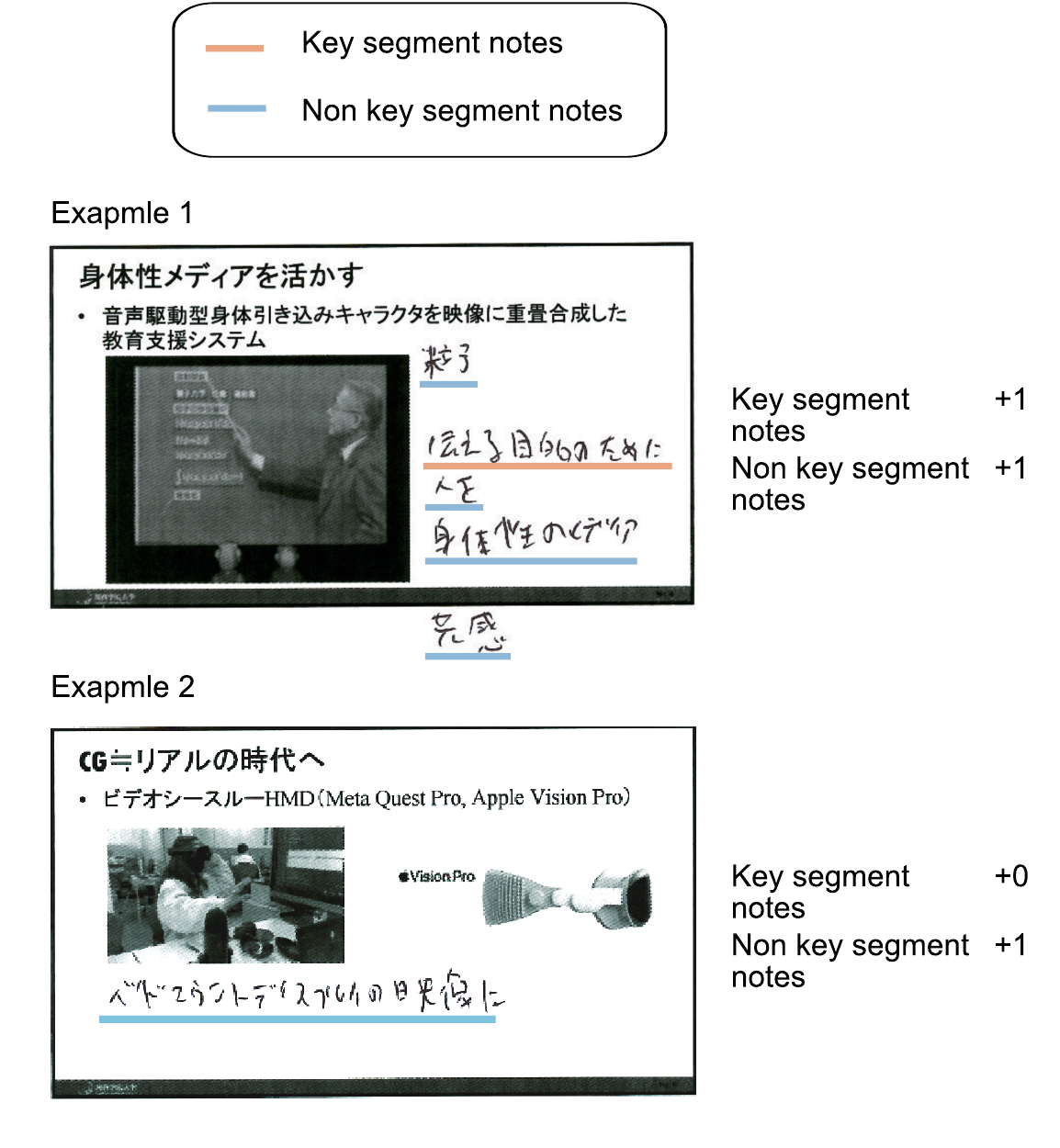}
		\end{center}
		\vspace{-0.8cm}
		\caption{Examples of counting the number of notes.}
		\label{fig:メモ数え方_詳細}
\end{figure}

\subsubsection{Score of mini-test:}
This was a set of 19 four-choice questions based on lecture content not printed on the handouts.

\subsubsection{Number of notes:}


We classified each note as ``key segment'' if it related to the topics tested in the mini-test, and ``non key segment'' otherwise. 
We counted each slide on which the participants wrote at least one note of a given type (Fig. \ref{fig:メモ数え方_詳細}).


\subsubsection{Posture (torso angle):}

\begin{figure}[t]
	\begin{center}
		\begin{tabular}{c}
			\begin{minipage}{0.5\hsize}
				\begin{center}
					\includegraphics[width= 0.6 \linewidth]{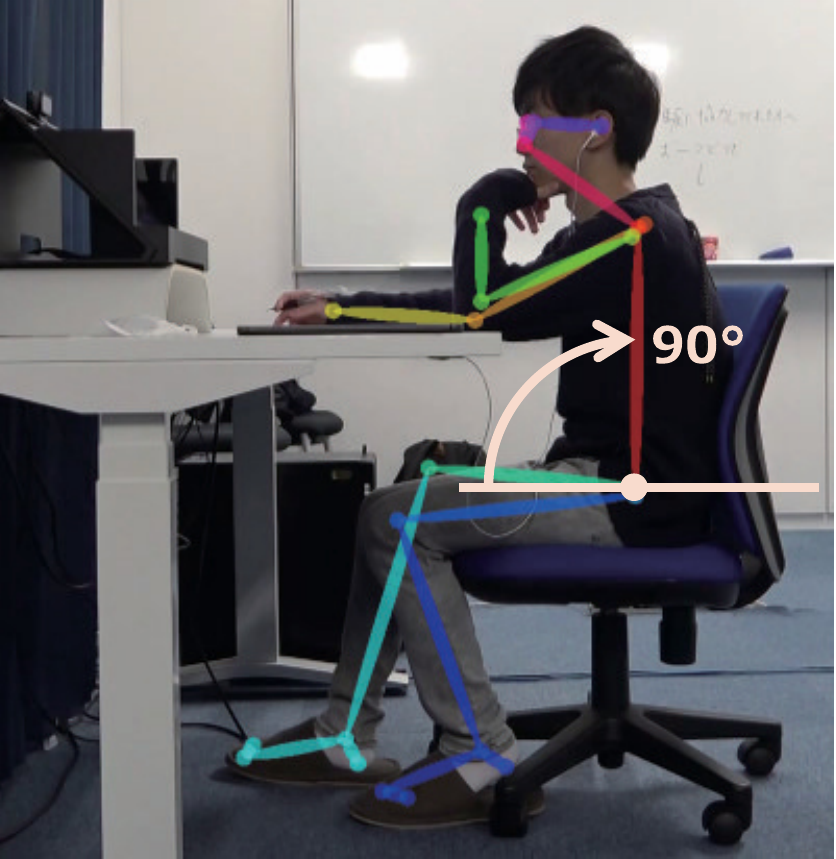}
				\end{center}
				
                \vspace{-0.6cm}
				\caption{$\vtop{\hbox{Posture }\hbox{(90 degrees).}}$}
				\label{fig:OpenPose_90}
			\end{minipage}
			\begin{minipage}{0.5\hsize}
				\begin{center}
					\includegraphics[width= 0.6 \linewidth]{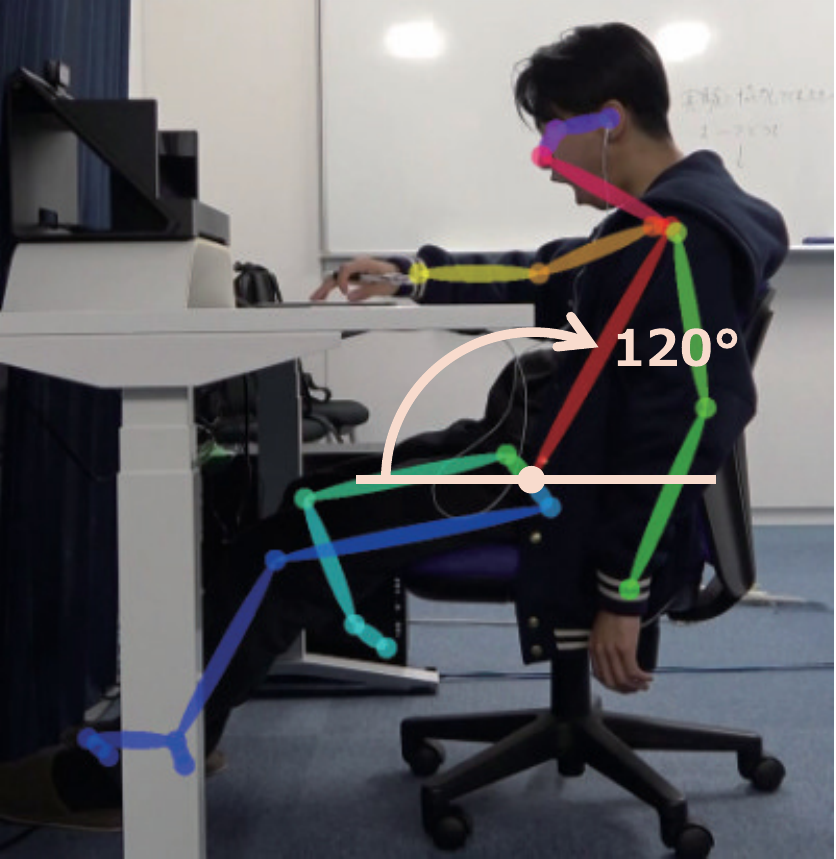}
				\end{center}
				
                \vspace{-0.6cm}
				\caption{$\vtop{\hbox{Posture }\hbox{(120 degrees).}}$}
				\label{fig:OpenPose_120}
			\end{minipage}
		\end{tabular}
	\end{center}
\end{figure}

The participants were video recorded, and their torso angles (Fig. \ref{fig:OpenPose_90}) were calculated using OpenPose v1.7.0 \cite{OpenPose}. Angles greater than 90 degrees indicated leaning back (Fig. \ref{fig:OpenPose_120}).

%

\section{Results}
\subsubsection{Questionnaire of impression:}

\begin{figure}[tbh]
		\begin{center}
			\includegraphics[width= 0.8 \linewidth]{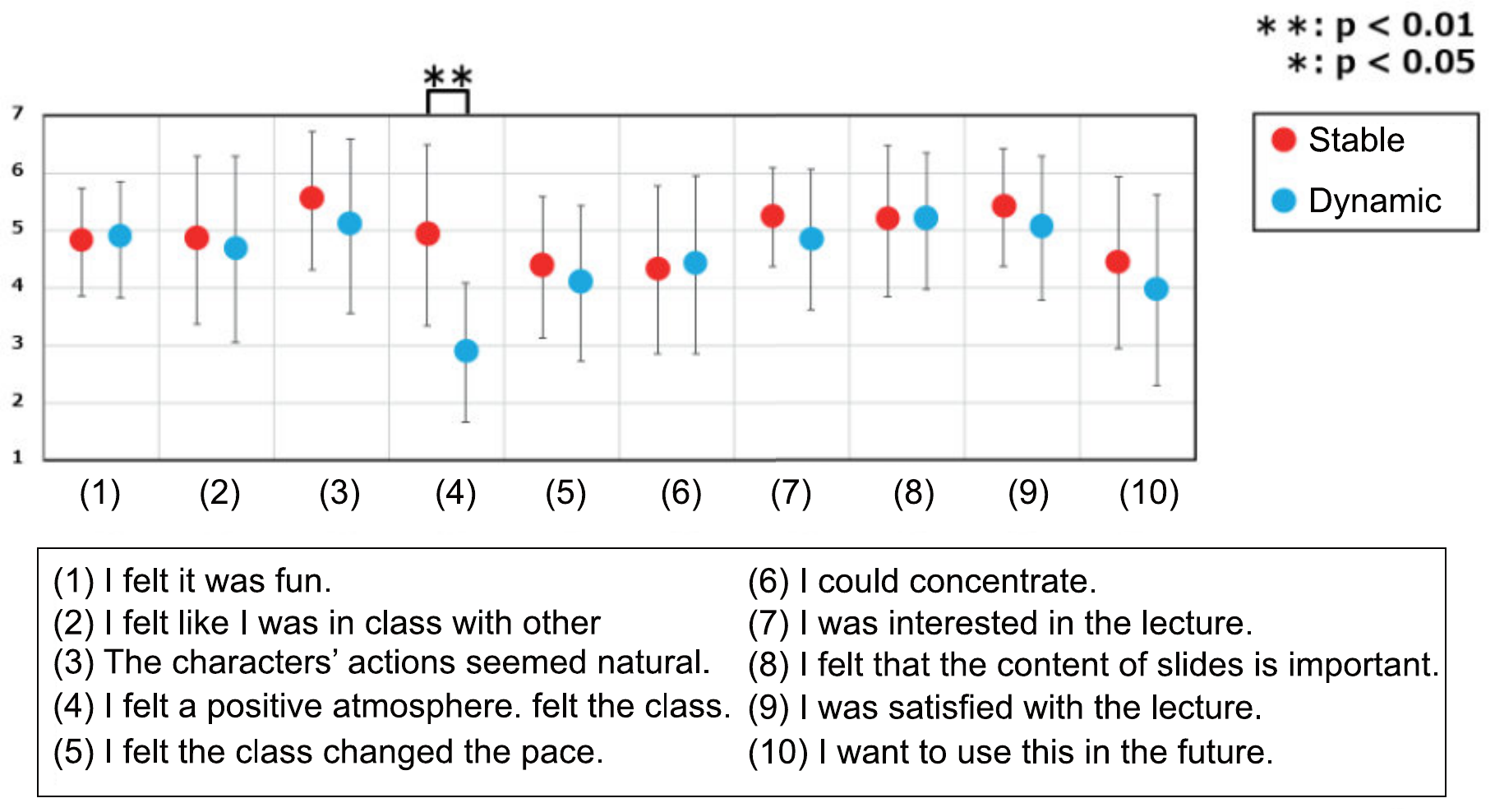}
		\end{center}
		\vspace{-0.6cm}
		\caption{Results of the impression  questionnaire.}
		\label{fig:授業アンケートの結果_30min}
\end{figure}

In general, the participants rated many items favorably, including ``I felt like I was in class with other students'' (Fig. \ref{fig:授業アンケートの結果_30min}). 
They also scored high on item 3 (``The characters' actions seemed natural'') for both modes, suggesting that our system preserved a sense of realism over the 33-minute span. 
A Wilcoxon rank sum test showed a significant difference between the two modes for item 4 (``I felt the atmosphere was positive''), reflecting that the inclusion of more negative actions in the dynamic mode led to a less positive impression.

\subsubsection{Score of mini-test:}

\begin{figure}[tbh]
		\begin{center}
			\includegraphics[width= 0.8 \linewidth]{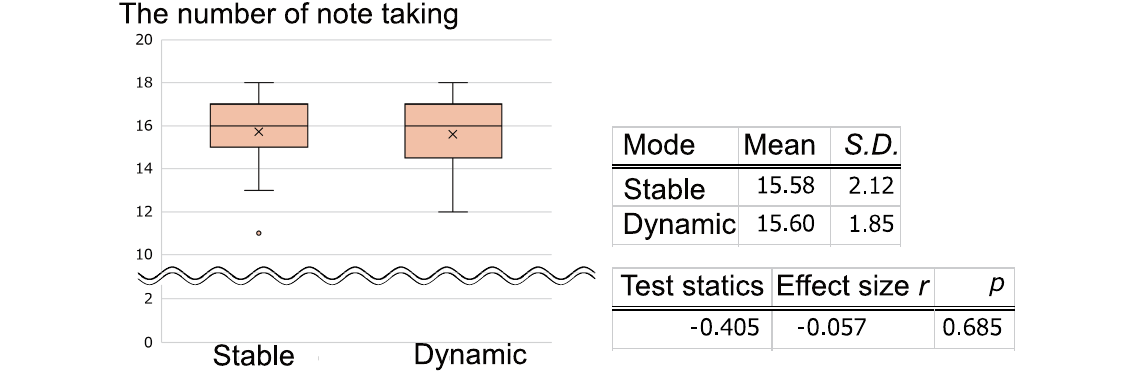}
		\end{center}
		\vspace{-0.3cm}
		\caption{Minitest scores.}
		\label{fig:授業成績_30min}
\end{figure}

A comparison of mini-test scores between Stable Mode and Dynamic Mode (Fig. \ref{fig:授業成績_30min}) revealed no statistically significant difference (Wilcoxon rank-sum test). 
The overall average was 15.7 out of 19 (~82.4\%), indicating that the test difficulty was suitable for a single class session.

\subsubsection{Number of notes:}

\begin{figure}[tbh]
		\begin{center}
            \includegraphics[width=  0.7 \linewidth]{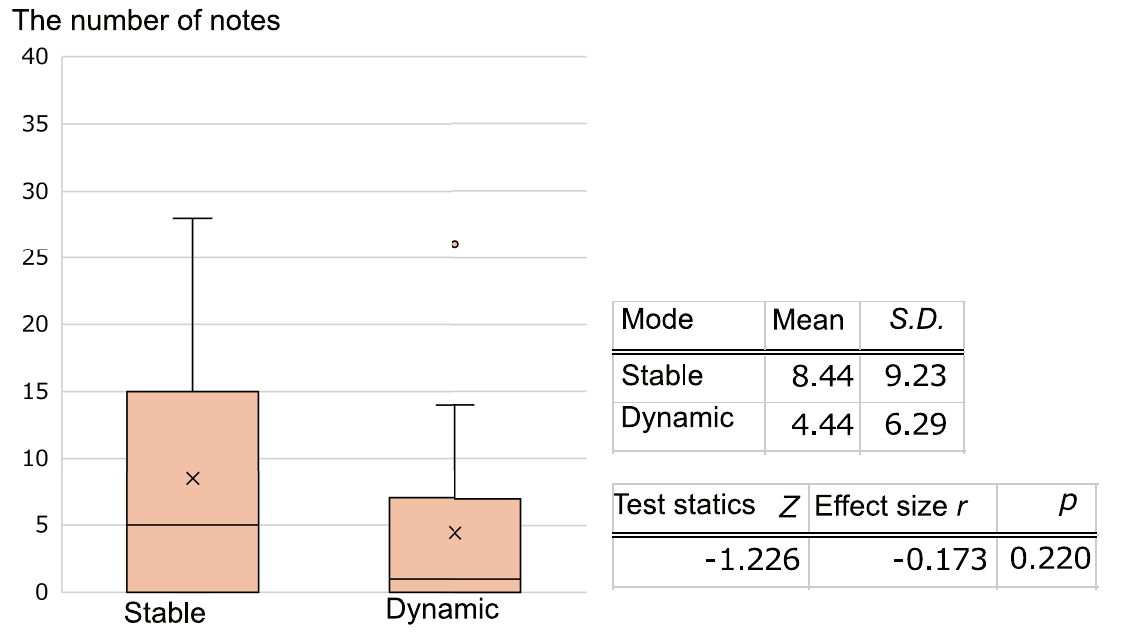}
		\end{center}
		\vspace{-0.4cm}
		\caption{Number of notes by all participants.}
		\label{fig:メモ数（P-PM）_全体}
\end{figure}

Initially, a Wilcoxon rank sum test found no significant differences in the total notes (Fig. \ref{fig:メモ数（P-PM）_全体}). 
However, 16 participants (8 in each mode) did not write any notes. 
To analyze the data in more detail, we conducted the test only with the students who took notes at least once, excluding them.

\begin{figure}[tbh]
		\begin{center}
            \includegraphics[width= 0.7 \linewidth]{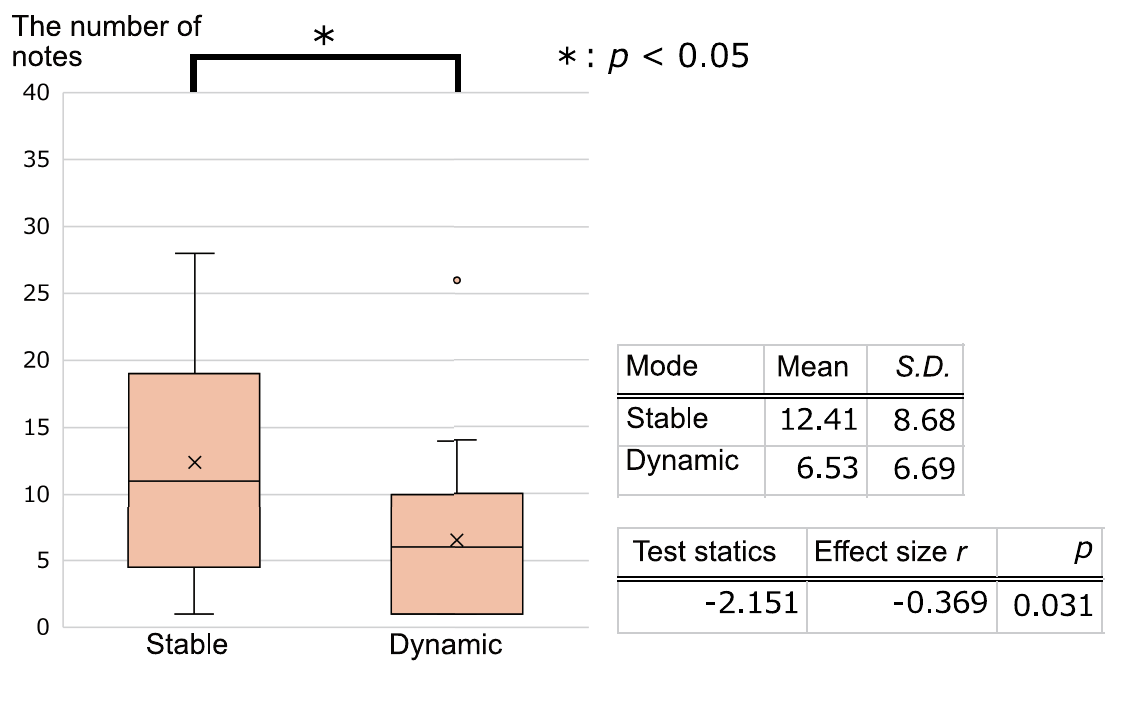}
		\end{center}
		\vspace{-0.6cm}
		\caption{Number of notes, excluding participants who did not take notes.}
		\label{fig:メモ数（P-PM）_メモなし除く}
\end{figure}

\begin{figure}[tbh]
		\begin{center}
                \includegraphics[width=  0.7 \linewidth]{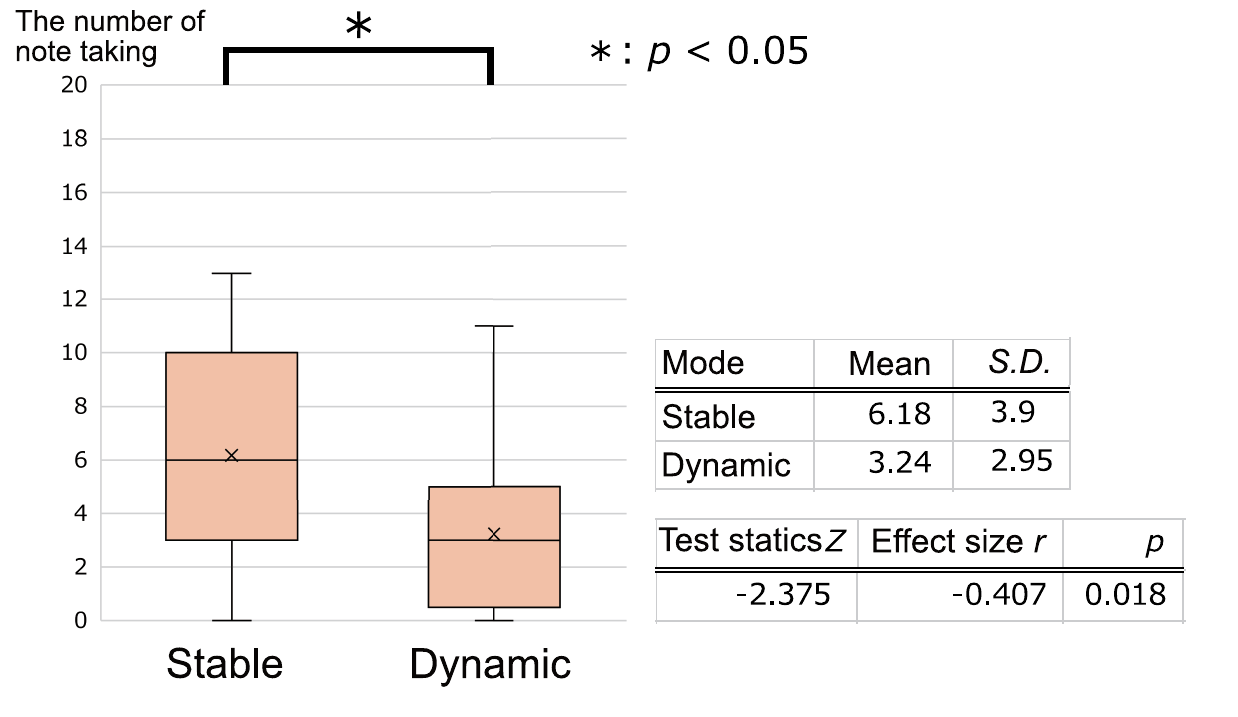}
		\end{center}
		\vspace{-0.4cm}
		\caption{Number of notes taken in key-segments, excluding participants who did not take notes.}
		\label{fig:メモ数_重要場面}
\end{figure}

\begin{figure}[tbh]
		\begin{center}
			\includegraphics[width=  0.7 \linewidth]{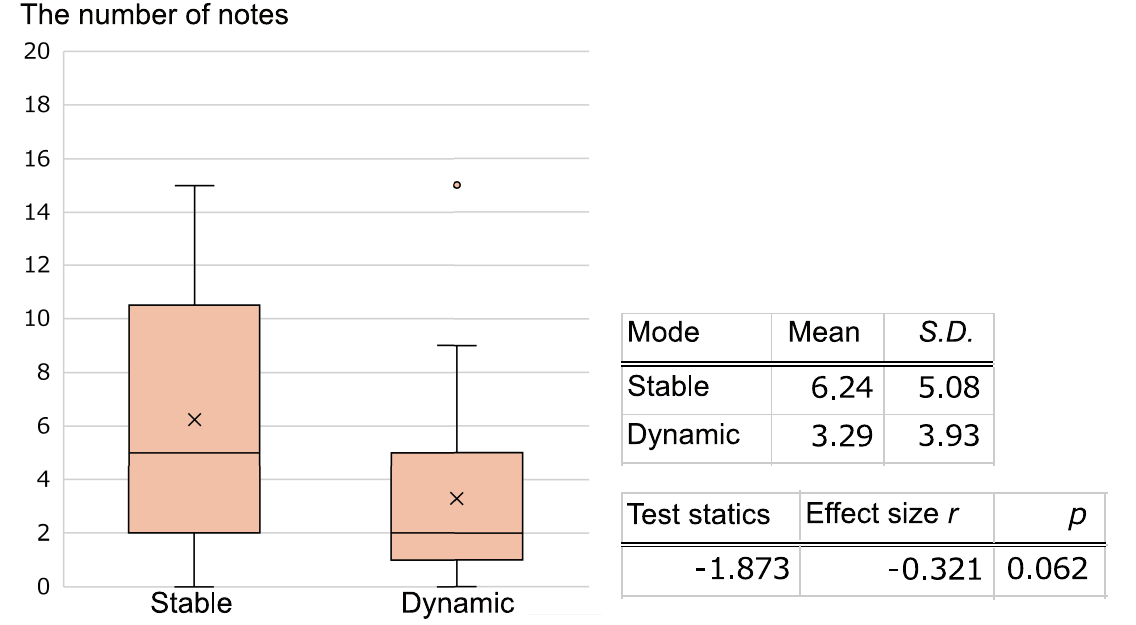}
		\end{center}
		\vspace{-0.4cm}
		\caption{Number of notes taken outside key-segments, excluding participants who did not take notes.}
		\label{fig:メモ数_重要場面外}
\end{figure}

The results of the test showed that participants in the Stable Mode group took significantly more notes (Fig. \ref{fig:メモ数（P-PM）_メモなし除く}). 
This gap was especially evident for key segment notes (Fig. \ref{fig:メモ数_重要場面}), whereas non-key segment notes showed no difference (Fig. \ref{fig:メモ数_重要場面外}).

\subsubsection{Mini-test score and number of notes:}
\begin{figure}[tbh]
		\begin{center}
            \includegraphics[width= 1 \linewidth]{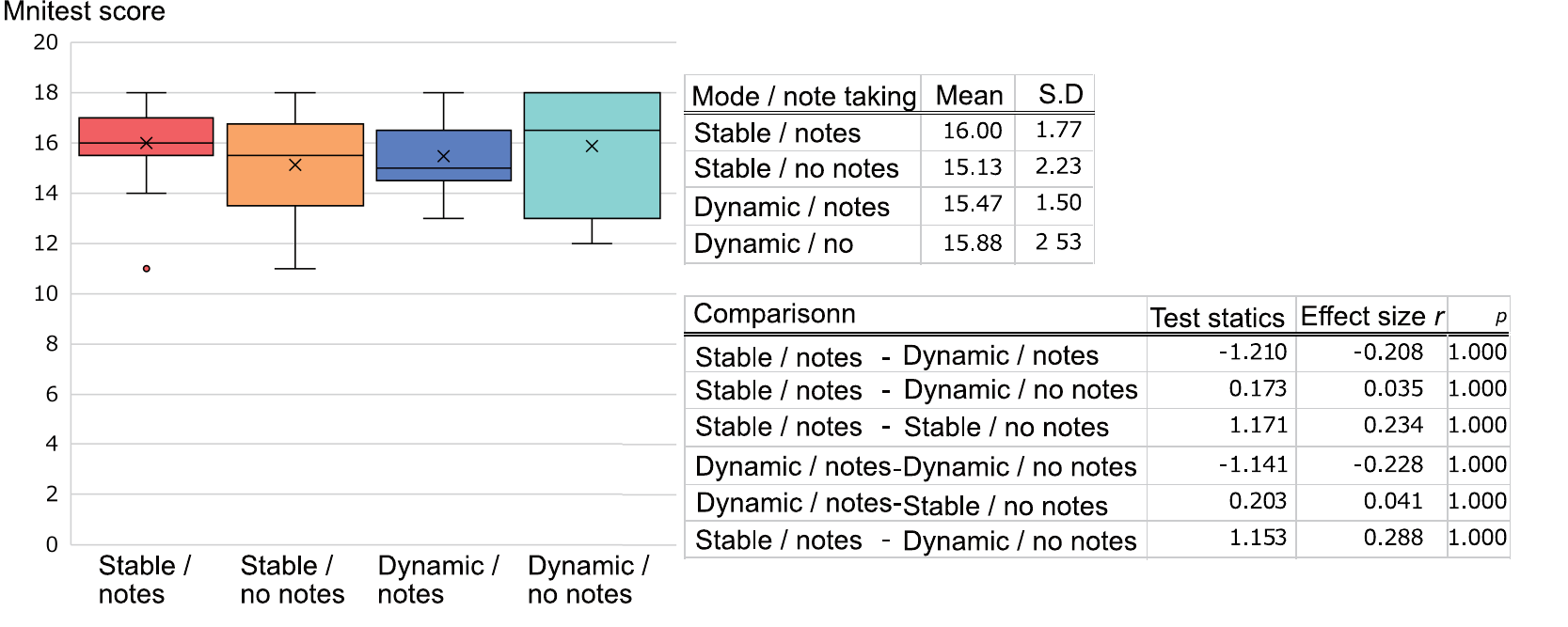}
		\end{center}
		\vspace{-0.4cm}
		\caption{Mini-test scores in each mode with and without notes.}
		\label{fig:メモ有無での成績}
\end{figure}

We divided the participants according to both mode and notetaking (Stable / notes, Stable / no notes, Dynamic / notes, Dynamic / no notes) and applied a Kruskal$–$Wallis test with Bonferroni corrections (Fig. \ref{fig:メモ有無での成績}).
We found no overall significant differences.

\begin{figure}[tbh]
		\begin{center}
            \includegraphics[width= 0.7 \linewidth]{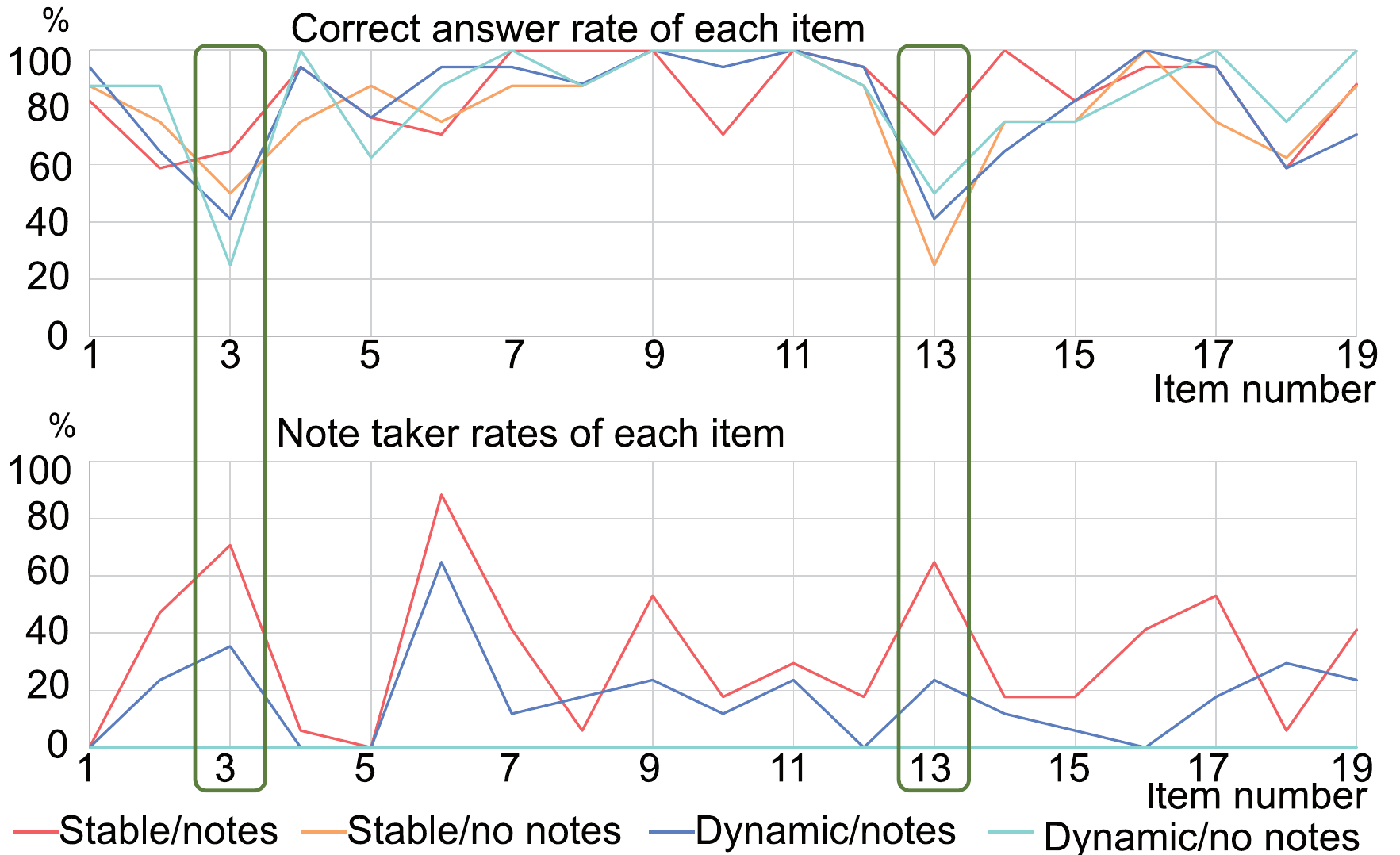}
		\end{center}
		\vspace{-0.6cm}
		\caption{Percentage of correct answers and percentage of those who took notes at each question.}
		\label{fig:正答率とメモ率}
\end{figure}

However, question-by-question analysis revealed that on more difficult items (lower average accuracy), notetakers in Stable Mode achieved notably higher scores (Fig. \ref{fig:正答率とメモ率}, green boxes). 
This suggests that taking notes in Stable Mode could be particularly helpful for comprehending challenging material.

\clearpage

\subsubsection{Posture (torso angle):}

\begin{figure}[tbh]
		\begin{center}
            \includegraphics[width= 0.8 \linewidth]{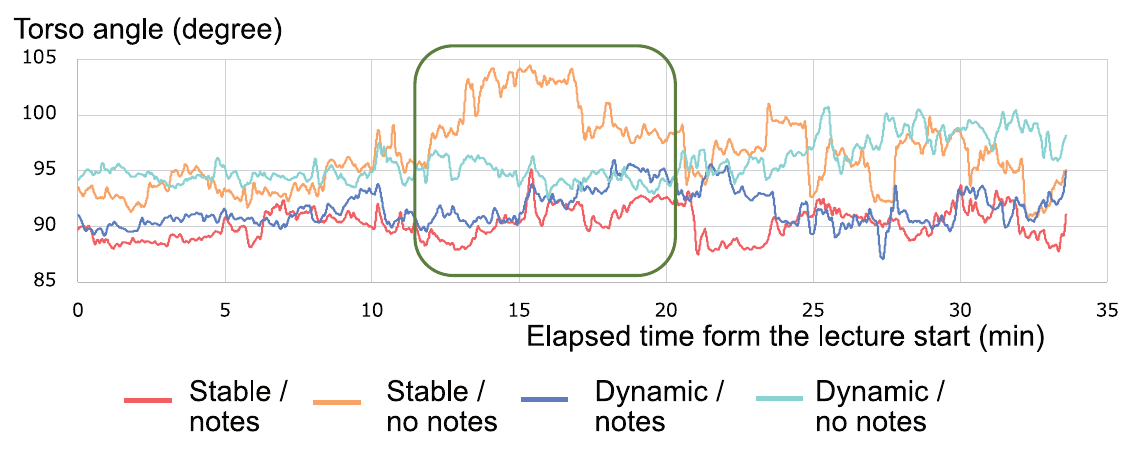}
		\end{center}
		\vspace{-0.4cm}
		\caption{Torso angles by mode and with/without notes (time-series graph of mean values).}
		\label{fig:姿勢_時系列}
\end{figure}

\begin{figure}[tbh]
		\begin{center}
        \includegraphics[width= 0.8 \linewidth]{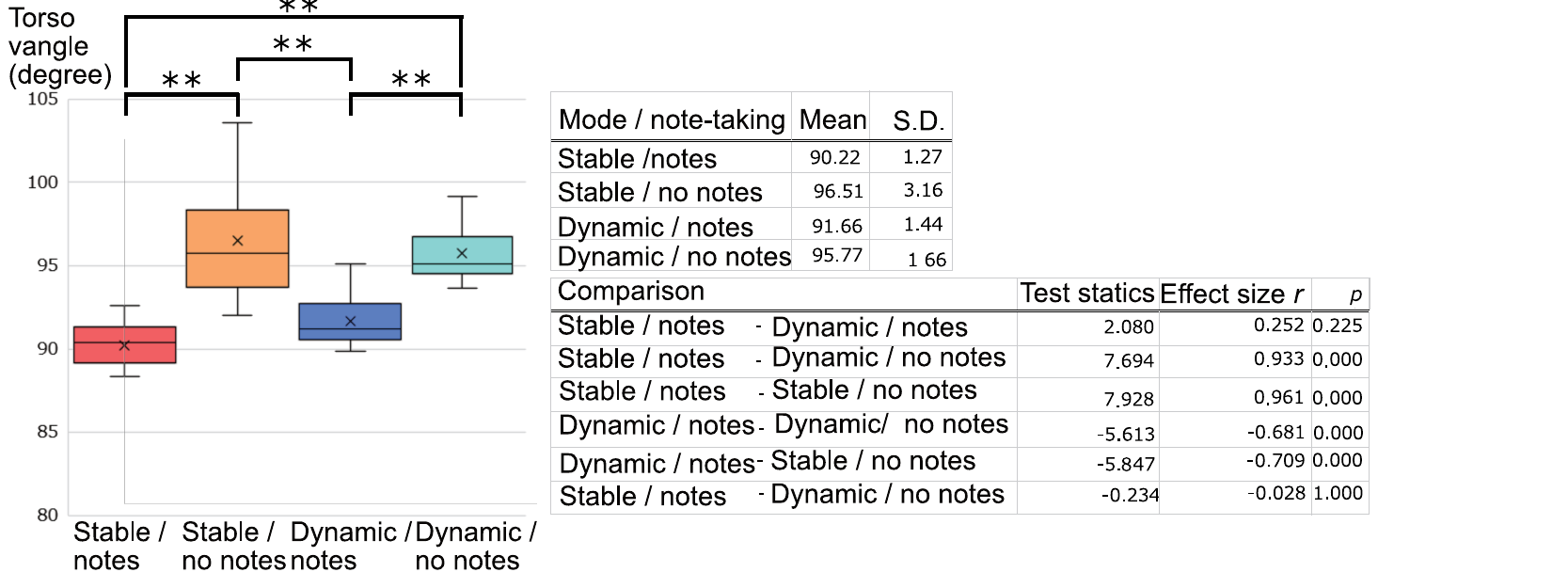}
		\end{center}
		\vspace{-0.4cm}
		\caption{Torso angles by mode and with/without notes (test results).}
		\label{fig:姿勢_検定}
\end{figure}

Figure \ref{fig:姿勢_時系列} displays the average posture angles over time for the four groups. 
We also took 1-minute averages and used a Kruskal$-$Wallis test with Bonferroni corrections (Fig. \ref{fig:姿勢_検定}).
In both modes, participants who took notes maintained a more upright posture, whereas those who did not leaned back more. 
In particular, between about 12 and 20 minutes (green box in Fig. \ref{fig:姿勢_時系列}), participants in Stable Mode who did not take notes showed marked backward leaning, likely indicating a decrease in engagement. 
However, participants in Dynamic Mode, regardless of notetaking behavior, legged less.

\section{Discussion}


The responses of the questionnaire show that most of the participants felt as if they were attending the class with other students, and they perceived characters' actions as natural throughout the 33-minute lecture. 
In addition, Stable Mode received higher scores in ``I felt the atmosphere was positive,'' according to its greater emphasis on positive actions.


Figures \ref{fig:メモ数（P-PM）_メモなし除く} and \ref{fig:メモ数_重要場面} indicate that Stable Mode also encouraged greater notetaking, particularly for key-segment notes, probably because the characters consistently showed nodding or notetaking. 
In contrast, although the Dynamic Mode also switched to positive actions at key segments, it did not match the Stable Mode in prompting notetaking.
This is contrary to our assumption.

The mini-test scores did not differ significantly by mode, but the participants who took notes in Stable Mode outperformed others on the more challenging questions (Fig. \ref{fig:授業成績_30min}). 
This suggests that if the lecture content were more difficult overall, the cues given by frequent positive actions in Stable Mode might be even more beneficial.


From Fig. \ref{fig:姿勢_時系列}, we see that participants in Stable Mode who did not take notes lean back more around the 12 to 20 minute mark, possibly because a low level of variety made them lose interest. 
In contrast, in Dynamic Mode, neither the group that took notes nor the group that did not take notes showed a backward-leaning posture, indicating that strategic shifts from negative to positive actions can help keep them physically alert. 
Consequently, Stable Mode seems more effective at encouraging notetaking, whereas Dynamic Mode helps prevent leaning back over long periods.



Combining a naked-eye 3D display with naturalistic character actions appears to successfully manage the communication field and impact learner focus in on-demand lectures. 
Although this study centered on on-demand classes, it may also enhance synchronous online learning if real students can appear as avatars alongside CG classmates, thus reinforcing copresence. 
A VR extension could heighten immersion even further, although care would be needed to ensure character motions look realistic at the life-size scale.

\section{Conclusion}


We developed an on-demand lecture watching system in which multiple student characters perform various actions, shaping a communication field. 
We tested two modes in an actual educational setting.
In Stable Mode, characters maintain positive actions (notetaking, nodding) for the most of the lecture. 
In Dynamic Mode, characters typically do negative actions, switching to positive ones only at key segments of the lecture video.
Our results show that Stable Mode promotes notetaking and helps participants score better on difficult questions, although some students who did not take notes lost interest. 
Meanwhile, Dynamic Mode reduced the tendency to lean back, but did not increase notetaking as effectively. 
As naked-eye 3D displays become increasingly common, our approach may prove practical in educational contexts. 
A VR-based version would likely deepen immersion, yet would require further study to ensure that character animations remain convincing at full scale.

\clearpage

\begin{credits}
This work was supported by JSPS KAKENHI Grant Number JP24K15641.

\subsubsection{\discintname}
The authors have no competing interests to declare that are relevant to the content of this article. 
\end{credits}

%
%
%
%

\end{document}